\documentstyle[aps,12pt]{revtex}

\begin{document}
\title{The existence of the fine electronic structure in {\it \ }LaCoO$_3$}
\author{Z. Ropka}
\address{Center for Solid State Physics, \'{s}w. Filip 5, 31-150 Krak\'{o}w,}
\author{R.J. Radwa\'{n}ski}
\address{Center for Solid State Physics, \'{s}w. Filip 5, 31-150 Krak\'{o}w; and\\
Inst. of Physics, Pedagogical University, 30-084 Krak\'{o}w, Poland.\\
email: sfradwan@cyf-kr.edu.pl.}
\maketitle

\begin{abstract}
We argue that in LaCoO$_3$ exists the fine electronic structure associated
with the atomic-like states of the Co$^{3+}$ ions and caused by the
crystal-field and intra-atomic spin-orbit interactions. This low-energy fine
electronic structure has to be taken into account for any meaningful
analysis of electronic and magnetic properties of LaCoO$_3$.

PACS\ No: 71.28 : 75.10.D : 71.70.Ej;


magnetic properties, LaCoO$_3$

Receipt date 10.02.1999 (Phys.Rev. B)
\end{abstract}

\date{(10.02.1999)}

Properties of LaCoO$_3$, a non-magnetic ground state and an anomalous
temperature dependence of the magnetic susceptibility ($\chi $), are still
intriguing despite of more than 30 years of intensive theoretical and
experimental studies. See, Refs 1-8 and the current literature in Phys.Rev.
B and Phys.Rev.Lett.

The aim of this short paper is to put attention that in LaCoO$_3$ the fine
electronic structure exists. This fine electronic (f-e) structure is caused
by the action of the crystal-field (CEF) and spin-orbit (s-o) interactions
on the Co$^{3+}$ ions.

The Co$^{3+}$ ion has 6 electrons in the unfilled 3d shell. They form the
highly-correlated electron system 3d$^6$, the ground term of which is
described$^{9,10}$ by S=2 and L=2 ($^5$D). Its 25-fold degeneration is
removed by the CEF and the s-o coupling. In case of the dominancy of the CEF
interactions over the s-o coupling, as is generally accepted for the 3d
ions, we obtain for the quasi-octahedral site closely lying 15 levels
originating from the $^5$T$_{2g}$ cubic subterm [10]. Other 10 levels
originating from the $^5$E$_g$ cubic subterm lie 2-3 eV above.

The existence of this fine electronic structure is generally neglected, see
e.g. Refs 4-6, despite that the above-mentioned knowledge about the
formation of the $^5$D term can be found in text books$^9$. This neglect of
the f-e structure is in fact related with the neglection in the literature
of the s-o coupling. We argue that the s-o coupling has to be taken into
account for any meaningful analysis of electronic and magnetic (e-m)
properties of compounds containing 3d ions like it is for rare-earth
compounds$^{11-13}$. This can be easily understood. The overall splitting of
the $^5$T$_{2g}$ ground subterm by the s-o coupling amounts approximately to
5$\lambda $. With $\lambda $ of 
\mbox{$\vert$}
20%
\mbox{$\vert$}
meV it means that there exists 15 discrete levels within the 100 meV range.
It yields an average energy separation of 7 meV. The calculated by us the
f-e structure of the Co$^{3+}$ ion in the slightly distorted octahedral
site, relevant to the situation realized in LaCoO$_3$, is shown in Fig. 1.
There is the non-magnetic singlet (in the $\left| \text{LSL}_{\text{z}}\text{%
S}_{\text{z}}\right\rangle $ space) ground state and two excited doublets
that turn out to be highly magnetic. These excited states become thermally
populated with the increasing temperature.

The existence of such discrete levels affects electronic and magnetic
properties, as we know well from rare-earth compounds$^{11-13}$. In
particular, some anomalies of the heat capacity (c) and of the magnetic
susceptibility occur at temperatures comparable with the first energy
separations. In the present case it is below 120 K ($\approx $ 11 meV). The
experimentally observed anomalies in $\chi $(T)$^8$ and c(T)$^1$ we are
taking as the confirmation of the presence of the fine-electronic structure
in LaCoO$_3$. The calculations$^{14}$, resembling those presented in Refs
11-13, reveal the Schottky-type maximum in the c(T) curve at about 60 K and
the rounded maximum in the $\chi (T)$ curve at 90 K. The obtainable
nonmagnetic ground state within the $^5$T$_{2g}$ subterm is very remarkable
result. This subterm has been up to now considered as the source of the
high-spin state only (compare please ref.15, p. 4258). The low-spin (i.e.
nonmagnetic) state has been attributed to the $^1$A$_1$ term (Refs 15, 4).
In contrary to the two-term consideration in the current literature, see
Refs 4 and 15, in our calculations we get the low- and high-spin state
within the one term (for it, the intra-atomic spin-orbit coupling is
essentially important). Moreover, in the energy level scheme shown in Fig. 1
one can also find the origin for the intermediate-spin state (the first and
the second excited states of Fig. 1 one can try to describe by the effective
spin of 1.16 and 1.83, respectively - these values are somehow close to the
ad hoc assumed values of S=1 and S=2 [Ref. 4]). These states are also in the
discussed energy interval (up to 80 meV). The possibility of getting the
non-magnetic ground state, discussed in the literature as the low-spin
state, as well as the intermediate and highly-magnetic states within the
same term we are taking as the great plus for our atomic-like approach.
Surely the explanation involving one term only is phsically simpler to be
realized. Moreover, according to the Occam's razor the simpler explanation
is the better one.

In conclusion, we argue that in LaCoO$_3$ exists the fine electronic
structure associated with the atomic-like states of the Co$^{3+}$ ions and
caused by the crystal-field (CEF) and intra-atomic spin-orbit (s-o)
interactions. This fine electronic structure has to be taken into account
for any meaningful analysis of electronic and magnetic properties of LaCoO$%
_3 $. Our approach provides in the very natural way the non-magnetic
low-temperature state (the 3d$^6$ highly-correlated system is a non-Kramers
system) and the insulating state in LaCoO$_3$. The present calculations
correct the electronic-structure considerations presented in Refs 1-8. In
particular, it turns out that the low-energy electronic structure is much
more complex than that considered in Ref. 2 (Fig. 1) and points out the
existence of the discrete states in contrary to the continuum energy-band
structure derived in Ref. 6 (Figs 1 and 2), in Ref. 4 (Figs. 1-3, 7) and in
Ref. 5.

The note added during the referee process (15.05.2000). 1. The present
discussion of LaCoO$_3$ differs significantly from the conventional one when
the low-spin nonmagnetic state $^1$A$_1$ is realized for strong enough
crystal fields. Our nonmagnetic state is found within the weak crystal-field
regime, i.e. when the crystal field does not break the intra-atomic
arrangement. 2. We would like to point out that our approach should not be
considered as the treatment of an isolated ion only - we consider the Co$%
^{3+}$ ion in the oxygen octahedron. The physical relevance of our treatment
to LaCoO$_3$ is obvious - the perovskite structure is built up from the
corner sharing Co$^{3+}$ octahedra. 3. We do not agree with the referee that
the splitting of only 11 meV can be easily exceeded by the exchange
interaction between Co ions leading to the magnetically-ordered ground state
of LaCoO$_3.$ As the confirmation of our approach we can mention that the
temperature dependence of the paramagnetic susceptibility resulting from our
approach [16] reproduces very well the one observed for LaCoO$_3$. 4. The
present approach, pointing out the existence of the discrete energy
spectrum, fundamentally differs from the continuum energy-band structure
considerations discussed recently in Ref. 17.

{\bf Acknowledgement}.The authors are grateful to R.Michalski for the
technical assistance.

{\bf Figure caption:}

Fig. 1. The lowest part of the fine electronic structure of the Co$^{3+}$
ion in LaCoO$_3$ originating from the cubic subterm $^5$T$_{2g}$. Other 10
states of the $^5$D term originating from the $^5$E$_g$ cubic subterm are
2.0-3.0 eV above - they practically do not influence the magnetic and
electronic properties of LaCoO$_3$. The non-magnetic ground state as well as
the intermediate and high-magnetic states should be noticed in this
atomic-like fine electronic structure.


\begin{references}
\bibitem{14}  the computer program can be obtained on the written request to
the authors
\end{references}
\end{document}